\documentclass[prl, twocolumn,amsmath,amssymb]{revtex4}
\usepackage{graphicx}
\usepackage{color}
\begin{document}

\title{Parametric Resonance of Optically Trapped Aerosols}
\author{R.~Di Leonardo$^1$}
\email[]{roberto.dileonardo@phys.uniroma1.it}
\author{G.~Ruocco$^1$}
\author{J.~Leach$^2$}
\author{M.~J.~Padgett$^2$}
\author{A.~J.~Wright$^3$}
\author{J.~M.~Girkin$^3$}
\author{D.~R.~Burnham$^4$}
\author{D.~McGloin$^4$}
\affiliation{
$^1$ INFM-CRS SOFT c/o Universit\'a di Roma ``La Sapienza'', I-00185, Roma, Italy.\\
$^2$ SUPA, Department of Physics \&  Astronomy, University of Glasgow, Glasgow, Scotland\\
$^3$ Institute of Photonics, SUPA, University of Strathclyde, Glasgow, Scotland\\
$^4$ SUPA, School of Physics \& Astronomy, University of St Andrews, St Andrews, Scotland\\
}
\date{\today}

\begin{abstract}

The Brownian dynamics of an optically trapped water droplet are investigated
across the transition from over to under-damped oscillations. The spectrum of
position fluctuations evolves from a Lorentzian shape typical of over-damped
systems (beads in liquid solvents), to a damped harmonic oscillator spectrum
showing a resonance peak.  In this later under-damped regime, we excite
parametric resonance by periodically modulating the trapping power at twice the
resonant frequency.  The power spectra of position fluctuations are in
excellent agreement with the obtained analytical solutions of a parametrically
modulated Langevin equation.

\end{abstract}

\maketitle



Parametric resonance provides an efficient and straightforward way to pump
energy into an underdamped harmonic oscillator \cite{broeck}. In general, if
the resonance frequency of an oscillator is dependent upon   a number of
parameters modulating any of these at twice the natural oscillation frequency
parametrically excites the resonance.  Such behavior leads to surprising
phenomena in the macroscopic world (pumping a swing, stability of vessels,
surface waves in vibrated fluids) \cite{ruby, bech}.  On the microscopic scale,
where stochastic forces become important, one refers to Brownian parametric
oscillators \cite{zerbe}. As an example, the parametric driving of Brownian
systems has been shown to be at the origin of some peculiar behaviors such as
the squeezing of thermal noise in Paul traps \cite{izmailov}. Parametrically
excited torsional oscillations have also been reported in a single-crystal
silicon microelectromechanical system \cite{Turner}.  What makes parametric
resonance useful is that in many cases it is easier to modulate a system
parameter rather than applying an oscillating driving force. Moreover, for
finite but low damping rates, we may never reach a stationary state with the
damping forces dissipating all of the input power and consequently the amplitude of
oscillations diverge. Optically trapped microparticles constitute a beautiful
example of Brownian damped harmonic oscillator (DHO) and they are becoming an
increasingly common tool for the investigation of different fields of basic and
applied science \cite{petrov}.  The possibility of pumping mechanical energy
into optically trapped particles could open the way to many applications.  In
optical tweezers, even though it is easy to periodically modulate the laser
power, parametric excitation is usually ineffective because of the heavy
damping action of the surrounding fluid.   

Recently it has been reported that modulating the laser
power at the parametric resonant frequency in an overdamped system increased the amplitude
of mean squared fluctuations  \cite{joykutty}.  However,
these findings have been difficult to reproduce
and are in strong contrast with the prediction of Langevin
dynamics \cite{pedersen, deng, deng2}.

In this Letter we show how parametric resonance can be excited in optically
trapped water droplets suspended in air, due to the reduced damping force.  We measure
power spectra of position fluctuations and find an excellent agreement with the
theoretical expectations based on Langevin dynamics with a parametric
forcing.

The dynamics of an optically trapped droplet is described by the Langevin
equation \cite{chandra}:

\begin{equation}
\label{langevin}
\ddot x(t)+\Omega_0^2 x(t)+\Gamma_0\dot x(t)=\xi(t)
\end{equation}

where $\Omega_0^2=k/m$ is the natural angular frequency of the
oscillator depending on trap stiffness $k$ and particle mass $m$.
$\Gamma_0=6\pi\eta a/C_c m$  is the
viscous damping due to the medium viscosity $\eta$ and depending on
particle radius $a$ and mass $m$. One must be careful
to consider the non-continuum effects of Stokes' Law in air due to
the finite Knudsen number of the particles under study. To correct
Stokes' Law the empirical slip correction factor, $C_c$, is
introduced, with a 5.5-1.6$\%$ reduction in drag for 3-10$\mu m$
diameter droplets respectively \cite{seinfeld}.

The stochastic force $\xi$, due to thermal agitation of solvent molecules,
is generally assumed to be uncorrelated on the time-scale of
particle's motion:

\begin{equation}
\label{noisecorr}
\langle\xi(0)\xi(t)\rangle=2 \Gamma_0 K_B T/m\;\delta(t)
\end{equation}

The corresponding power spectrum $S_\xi(\omega)$ of the stochastic variable
$\xi$ can be defined as:

\begin{equation}
\label{noise}
\langle\hat\xi^*(\omega)\hat\xi(\omega')\rangle=
S_\xi(\omega)
\delta(\omega-\omega')
\end{equation}

where:

\begin{equation}
\label{ft}
\hat\xi(\omega)=1/2\pi\int_{-\infty}^{\infty}\xi(t)\exp[-i\omega t] dt
\end{equation}

By putting (\ref{ft}) and (\ref{noisecorr}) in  (\ref{noise}) it is easily shown
that $\xi$ has a white noise spectrum
\begin{equation}
\label{noisespectrum}
S_\xi(\omega)=\frac{\Gamma_0 K_B T}{\pi m}
\end{equation}

By Fourier transforming (\ref{langevin}) and using (\ref{noisespectrum}) we can easily
obtain the power spectrum of position fluctuations as:

\begin{equation}
\label{dho}
S_x(\omega)=\frac{K_B T}{k}
\frac{1}{\pi}\frac{\Omega_0^2\; \Gamma_0}{(\omega^2-\Omega_0^2)^2+\Gamma_0^2\omega^2}
\end{equation}

When trapping objects in liquid solvents the ratio
$\Gamma_0/\Omega_0$ is always larger than one, i.e the system is
overdamped.  The ratio $\Gamma_0/\Omega_0$ 
depends only slightly on particle's size and, in solvents with water-like
viscosities, is always greater than 1 up to power levels of some tens of
Ws.  For typical trapping powers of order $10$ mW in water
$\Gamma_0/\Omega_0$ is typically $>10$.  As a result only those
frequencies smaller than $\Gamma_0$, and hence much smaller than
$\Omega_0$, have a significant amplitude in the power spectrum of
$x$. Under these conditions we can therefore neglect $\omega^2$ with
respect to $\Omega_0^2$ in the first term of the denominator in
(\ref{dho}) and obtain the usual Lorentzian power spectrum
characterised by $\omega^{-2}$ tails \cite{wang}.  Such an
overdamped condition precludes the possibility of exciting
significant oscillations either directly or parametrically.  To
probe oscillations in the liquid damped regime we would need to be
able to increase typical trap power by four orders of magnitude -
introducing uncontrollable heating and damage of the trapped object.
A more feasible route is to reduce viscosity by two orders of
magnitude. This last condition  can be readily obtained by trapping
particles in air whose viscosity is approximately
1/55$^{th}$ of water ($\eta=1.8\times 10^{-5}$ Pa s) \cite{seinfeld}.

For these experiments, our optical tweezers are based around an
inverted microscope with a high numerical  aperture oil immersion
microscope objective (1.3NA, $100\times$).  The continuous wave laser is
a Nd:YAG, frequency doubled to give 0-2 W of  532 nm light.  To
couple the beam into the air medium, a single cover slip is rested
over the objective on a thin oil layer.    A water aerosol is
produced using a nebulizer and usually a 3-10$\mu$m diameter water
droplet is trapped at the beam focus \cite{hopkins, mcgloin}, see
Fig.~1. One should  note that whereas for particles trapped in
fluid, a laser power of 10s of mW is typical, here, to maximise the
stiffness of the trap, we use powers of order 1 W.

A quadrant photo detector, placed in the back focal plane of the condenser
lens, receives the light transmitted through the droplet. By measuring the
imbalance of the light collected by the quadrants, the lateral displacement of
the droplet is deduced with a bandwidth of several kHz and a precision of better
than 5 nm \cite{gittes}.

\begin{figure} \includegraphics[width=.4\textwidth]{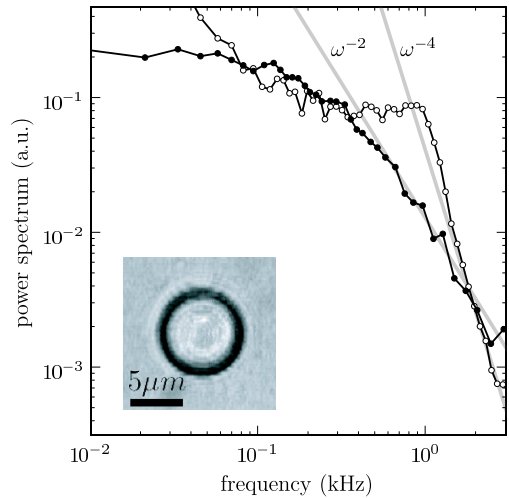} \caption{The
measured power spectra of trapped aerosol particle at two different powers.  At lower
power (black circles) it is overdamped and the mean squared amplitude of the high frequency
motion decays as $\omega^{-2}$.  At higher powers (white circles), the aerosol is underdamped
and the mean squared amplitude decays as $\omega^{-4}$.  The inset shows an
optical image of a trapped aerosol particle. }\label{transition} \end{figure}

The power spectra of the measured displacement, for two different trap powers, are
shown in Fig.  \ref{transition}.  It is clearly visible how the particle
dynamics crosses over from an over damped dynamics with a Lorentzian spectrum
with a high frequency roll-off proportional to $\omega^{-2}$ to an underdamped
regime with a faster roll-off, $\omega^{-4}$, and the appearance of a resonance
peak at a frequency of about 1 kHz. The emergence  of such a peak arises from
the fact that the inertial terms in (\ref{langevin}) are no longer negligible.
As a consequence an average trajectory starting away from the equilibrium position
crosses the equilibrium position with a finite velocity.

In this situation the parametric resonance is excited by modulating the
strength  of the trapping potential. Ideally the potential is made shallower
when the particle traverses the equilibrium position and steeper again when the
particle is far from the equilibrium position. This is maximally efficient when
we modulate the potential at twice the natural oscillation frequency $\Omega_0$.
To consider this  model in quantitative terms we can rewrite
(\ref{langevin}) in the presence of a parametrically modulated external
potential:

\begin{eqnarray}
\label{langepara}
&&\ddot x(t)+\Omega_0^2[1+g f(t)]x(t)+\Gamma_0\dot x(t)=\xi(t)\\
&&f(t+{\cal T})=f(t),\;\;-1<f(t)<1
\end{eqnarray}

where $0<g<1$ measures the strength of modulation. By Fourier transforming (\ref{langepara}) we obtain:

\begin{equation}
\label{fourierlp}
(-\omega^2+\Omega_0^2+i\omega\Gamma_0)\hat x(\omega)+\Omega_0^2 g \sum_{k=-\infty}^\infty a_k
\hat x(\omega+k\Omega_1)=\hat \xi(\omega)\\
\end{equation}

where $a_k$ is the coefficient of the $k 2 \pi/{\cal T}=k\Omega_1$ frequency
component of the Fourier series expansion of $f(t)$.  It is clear from equation 
(\ref{fourierlp}) how parametric modulation introduces a coupling between all
those frequencies differing by an integer number of $\Omega_1$. We now
introduce the vectors $X_n(\omega)=\hat x(\omega+n\Omega_1)$ and
$R_n(\omega)=\hat \xi(\omega+n\Omega_1)$ and write the recursive relations:

\begin{eqnarray}
\nonumber
[-(\omega+n\Omega_1)^2+\Omega_0^2+i(\omega+n\Omega_1)\Gamma_0]X_n(\omega)+\\
\label{fourierX}
\Omega_0^2 g \sum_{k=-\infty}^\infty a_k X_{n+k}(\omega)=R_n.
\end{eqnarray}

To obtain the power spectrum $S_x(\omega)$, for each frequency $\omega$ we
should compute $X_0(\omega)$.  This will be, in turn, coupled to all other
components in the array $X_n$.  However the strength of the coupling will decay
for large $|n|$ so that we can limit ourselves to a finite number of components
and write the matrix equation for the array

${\mathbf X}(\omega)=[X_{-N}(\omega),...,X_N(\omega)]$:

\begin{equation}
{\mathbf G}^{-1}(\omega) {\mathbf X}(\omega)={\mathbf R}(\omega)
\end{equation}

with
\begin{equation}
G^{-1}_{nk}(\omega)=
[-(\omega+n\Omega_1)^2+\Omega_0^2+i(\omega+n\Omega_1)\Gamma_0]\delta_{nk}+
\Omega_0^2 g  a_{k-n}
\end{equation}

By matrix inversion we obtain the power spectrum as:

\begin{eqnarray}
\nonumber
\langle X_0^*(\omega) X_0(\omega')\rangle&=&\sum_{k,n=-N}^{N}G_{0k}^*(\omega)G_{0l}(\omega')
\langle R_k^*(\omega)R_n(\omega')\rangle\\
&=&\frac{\Gamma_0 K_B T}{\pi m}\sum_{n=-N}^{N}|G_{0n}(\omega)|^2 \delta(\omega-\omega')
\end{eqnarray}

and from the definition of power spectrum:

\begin{equation}
\label{paraspectrum}
S_x(\omega)=\frac{\Gamma_0 K_B T}{\pi m}\sum_{k=-N}^{N}|G_{0k}(\omega)|^2
\end{equation}

\begin{figure}
\includegraphics[width=.45\textwidth]{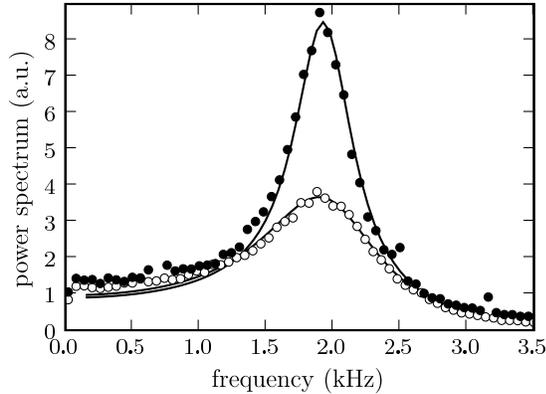}

\caption{The measure power spectrum of a trapped water droplet for no modulation of the
laser power (white circles) and modulation at 3.9 kHz ($\Omega_1\simeq 2 \Omega_0$) (black
circles).  The peak is higher and narrower on the resonant condition thus
indicating parametric excitation. Solid line below black circles is the predicted spectrum
from (\ref{paraspectrum}).}\label{onoff}

\end{figure}

If $\Omega_0$ and $\Gamma_0$ are known, we can use
(\ref{paraspectrum}) to predict the power spectrum of a Brownian
particle in a modulated trap.  The white circles of Fig.\ref{onoff}
show the measured power spectrum for a trapped water droplet when
the trap has constant power. The presence of the peak suggests that
we are in an underdamped regime.  By fitting to equation
(\ref{dho}) (solid line) we can directly extract the resonant
frequency $\Omega_0/2\pi=2.0$ kHz and the damping term
$\Gamma_0=6.8$ kHz relevant to our experimental conditions. 
The fitted value of $\Gamma_0$ corresponds to the
Stokes drag on an aerosol droplet of radius
3.4$\mu$m.

We then apply the square-wave modulation of the trapping power, with the
laser adjusted to give the same average power as before, $\Omega_1 \simeq 2
\Omega_0$ and $g=0.4$.  The black circles of Fig.\ref{onoff} show the power
spectrum of the lateral motion for the trapped water droplet with modulated laser power. The excitation of
a resonance appears as a higher and narrower peak in the power spectrum.  This
observed spectrum matches closely the expected behavior (solid line) obtained by applying
the measured parameters $\Omega_0$, $\Gamma_0$, $\Omega_1$, $g$ to equation
(\ref{paraspectrum}), strongly supporting our interpretation of this peak being
due to a parametric excitation of the resonance.
Higher harmonics, characterizing the response of parametrically driven systems,
are suppressed in our case, being only slightly underdamped.

\begin{figure} 
\includegraphics[width=.4\textwidth]{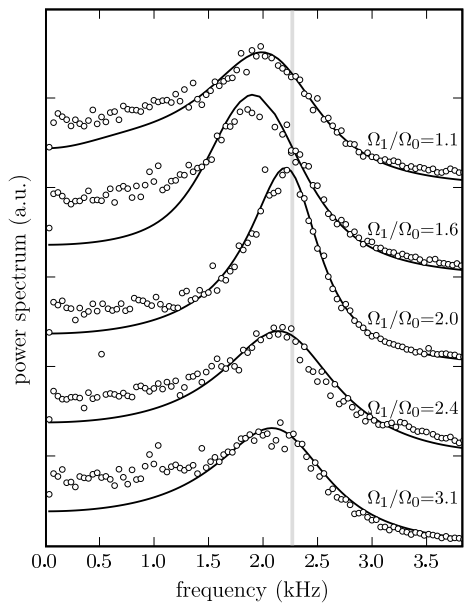}

\caption{Evolution of position power spectra on varying the modulation
frequency $\Omega_1$. Parametric excitation of oscillations is evident at the
parametric resonance condition $\Omega_1/\Omega_0=2$.  The solid lines are the
theoretical predictions from (\ref{paraspectrum}).}

\label{scan} \end{figure}

\begin{figure}
\includegraphics[width=.4\textwidth]{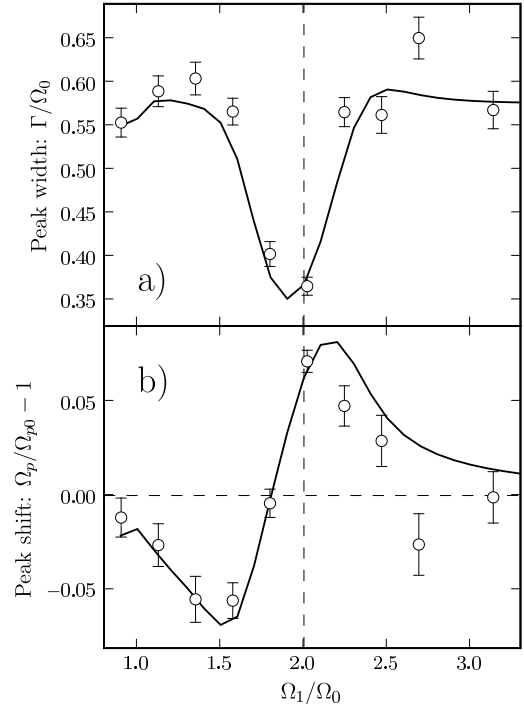}
\caption{
a) Peak fullwidth half maximum as a function of modulation frequency.
Solid lines are the theoretical predictions from (\ref{paraspectrum}).
b) Peak position shift as a function of modulation frequency.
}\label{gammaomega}
\end{figure}

Using these parameters with (\ref{paraspectrum}) we can make
general predictions about the dynamics of a particle that can then
be verified against experiments. One comparison to make is the
predicted and observed form of the power spectra as a function of
the modulation frequency, both above and below the parametric
resonance condition. These results are shown in Fig.~\ref{scan}, the
white circles are the experimental points and the black lines are
the predicted spectra.  Again, there is an excellent agreement
between the observed and predicted particle motion.  In particular
the parametric excitation of oscillations manifests as a narrowing
of the peak (or a reduced apparent damping $\Gamma$, defined as the
full width half maximum), occurring when modulating at twice
$\Omega_0$. A shift in peak position $\Omega_p$ is also apparent
close to parametric resonance. Both of these signatures are compared
with theoretical expectations in Fig. \ref{gammaomega} further
supporting our interpretation of the system as being a Brownian
parametric oscillator .

We recognise that the system we report here relates to study of only the
lateral motion of the trapped droplets.  In keeping with other work we note
from examination of the video images that the axial movement of the trapped
droplet has significant amplitude on much longer timescales,  corresponding to
frequencies in the region of 10-50Hz \footnote{This slow
axial dynamics is responsible for the low frequency component of spectra in
fig. \ref{gammaomega}}.  This reflects the comparatively weak axial trapping,
possibly arising from aberrations  associated with non optimised objectives or
the increased scattering force.  It may be possible to use doughnut or
Laguerre- Gaussian beams having zero on-axis intensity, and improved axial
trapping \cite{oneill}.

We have reported the first observation of a parametrically excited resonance
within a Brownian oscillator. The demonstration of this effect within optical
tweezers for this purpose was made possible by relying on  the viscosity of air
to lightly damp the motion of a trapped aerosol droplet.  The detailed observed
dynamics match closely the power spectra predicted from a parametrically
modulated Langevin equation, where all parameters were fixed  from examination
of the non modulated system.

We wish to thank S. Ciuchi for many helpful discussions.  This work was supported by the EPSRC.


\bibliographystyle{apsrev}

\begin{thebibliography}{18}
\expandafter\ifx\csname natexlab\endcsname\relax\def\natexlab#1{#1}\fi
\expandafter\ifx\csname bibnamefont\endcsname\relax
  \def\bibnamefont#1{#1}\fi
\expandafter\ifx\csname bibfnamefont\endcsname\relax
  \def\bibfnamefont#1{#1}\fi
\expandafter\ifx\csname citenamefont\endcsname\relax
  \def\citenamefont#1{#1}\fi
\expandafter\ifx\csname url\endcsname\relax
  \def\url#1{\texttt{#1}}\fi
\expandafter\ifx\csname urlprefix\endcsname\relax\def\urlprefix{URL }\fi
\providecommand{\bibinfo}[2]{#2}
\providecommand{\eprint}[2][]{\url{#2}}

\bibitem[{\citenamefont{den Broeck and Bena}(2000)}]{broeck}
\bibinfo{author}{\bibfnamefont{C.~V.} \bibnamefont{den Broeck}}
  \bibnamefont{and} \bibinfo{author}{\bibfnamefont{I.}~\bibnamefont{Bena}}
  (\bibinfo{year}{2000}).

\bibitem[{\citenamefont{Ruby}(1996)}]{ruby}
\bibinfo{author}{\bibfnamefont{L.}~\bibnamefont{Ruby}}, \bibinfo{journal}{Am.
  J. Phys.} \textbf{\bibinfo{volume}{64}}, \bibinfo{pages}{39}
  (\bibinfo{year}{1996}).

\bibitem[{\citenamefont{Bechhoefer and Johnson}(1996)}]{bech}
\bibinfo{author}{\bibfnamefont{J.}~\bibnamefont{Bechhoefer}} \bibnamefont{and}
  \bibinfo{author}{\bibfnamefont{B.}~\bibnamefont{Johnson}},
  \bibinfo{journal}{Am. J. Phys.} \textbf{\bibinfo{volume}{64}},
  \bibinfo{pages}{1482} (\bibinfo{year}{1996}).

\bibitem[{\citenamefont{Zerbe et~al.}(1994)\citenamefont{Zerbe, Jung, and
  H\"anggi}}]{zerbe}
\bibinfo{author}{\bibfnamefont{C.}~\bibnamefont{Zerbe}},
  \bibinfo{author}{\bibfnamefont{P.}~\bibnamefont{Jung}}, \bibnamefont{and}
  \bibinfo{author}{\bibfnamefont{P.}~\bibnamefont{H\"anggi}},
  \bibinfo{journal}{Phys. Rev. E} \textbf{\bibinfo{volume}{49}}
  (\bibinfo{year}{1994}).

\bibitem[{\citenamefont{Izmailov et~al.}(1995)\citenamefont{Izmailov, Arnold,
  Holler, and Myerson}}]{izmailov}
\bibinfo{author}{\bibfnamefont{A.~F.} \bibnamefont{Izmailov}},
  \bibinfo{author}{\bibfnamefont{S.}~\bibnamefont{Arnold}},
  \bibinfo{author}{\bibfnamefont{S.}~\bibnamefont{Holler}}, \bibnamefont{and}
  \bibinfo{author}{\bibfnamefont{A.~S.} \bibnamefont{Myerson}},
  \bibinfo{journal}{Phys. Rev. E} \textbf{\bibinfo{volume}{52}},
  \bibinfo{pages}{1325} (\bibinfo{year}{1995}).

\bibitem[{\citenamefont{Turner et~al.}(1998)\citenamefont{Turner, Miller,
  Hartwell, MacDonald, Strogatz, and Adams}}]{Turner}
\bibinfo{author}{\bibfnamefont{L.}~\bibnamefont{Turner}},
  \bibinfo{author}{\bibfnamefont{S.}~\bibnamefont{Miller}},
  \bibinfo{author}{\bibfnamefont{P.}~\bibnamefont{Hartwell}},
  \bibinfo{author}{\bibfnamefont{N.}~\bibnamefont{MacDonald}},
  \bibinfo{author}{\bibfnamefont{S.}~\bibnamefont{Strogatz}}, \bibnamefont{and}
  \bibinfo{author}{\bibfnamefont{S.}~\bibnamefont{Adams}},
  \bibinfo{journal}{Nature} \textbf{\bibinfo{volume}{396}},
  \bibinfo{pages}{149} (\bibinfo{year}{1998}).

\bibitem[{\citenamefont{Volpe and Petrov}(2006)}]{petrov}
\bibinfo{author}{\bibnamefont{Volpe}} \bibnamefont{and}
  \bibinfo{author}{\bibnamefont{Petrov}}, \bibinfo{journal}{Phys. Rev. Lett.}
  \textbf{\bibinfo{volume}{97}} (\bibinfo{year}{2006}).

\bibitem[{\citenamefont{Joykutty et~al.}(2005)\citenamefont{Joykutty, Mathur,
  Venkataraman, and Natarajan}}]{joykutty}
\bibinfo{author}{\bibfnamefont{J.}~\bibnamefont{Joykutty}},
  \bibinfo{author}{\bibfnamefont{V.}~\bibnamefont{Mathur}},
  \bibinfo{author}{\bibfnamefont{V.}~\bibnamefont{Venkataraman}},
  \bibnamefont{and}
  \bibinfo{author}{\bibfnamefont{V.}~\bibnamefont{Natarajan}},
  \bibinfo{journal}{Phys. Rev. Lett.} \textbf{\bibinfo{volume}{95}},
  \bibinfo{eid}{193902} (\bibinfo{year}{2005}).

\bibitem[{\citenamefont{Pedersen and Flyvbjerg}(in press)}]{pedersen}
\bibinfo{author}{\bibfnamefont{L.}~\bibnamefont{Pedersen}} \bibnamefont{and}
  \bibinfo{author}{\bibfnamefont{H.}~\bibnamefont{Flyvbjerg}},
  \bibinfo{journal}{Phys. Rev. Lett.}  (\bibinfo{year}{in press}).

\bibitem[{\citenamefont{Deng et~al.}(in press{\natexlab{a}})\citenamefont{Deng,
  Forde, and Bechhoefer}}]{deng}
\bibinfo{author}{\bibfnamefont{Y.}~\bibnamefont{Deng}},
  \bibinfo{author}{\bibfnamefont{N.~R.} \bibnamefont{Forde}}, \bibnamefont{and}
  \bibinfo{author}{\bibfnamefont{J.}~\bibnamefont{Bechhoefer}},
  \bibinfo{journal}{Phys. Rev. Lett.}  (\bibinfo{year}{in
  press}{\natexlab{a}}).

\bibitem[{\citenamefont{Deng et~al.}(in press{\natexlab{b}})\citenamefont{Deng,
  Bechhoefer, and Forde}}]{deng2}
\bibinfo{author}{\bibfnamefont{Y.}~\bibnamefont{Deng}},
  \bibinfo{author}{\bibfnamefont{J.}~\bibnamefont{Bechhoefer}},
  \bibnamefont{and} \bibinfo{author}{\bibfnamefont{N.~R.} \bibnamefont{Forde}},
  \bibinfo{journal}{J. Opt. A: Pure Appl. Opt.}  (\bibinfo{year}{in
  press}{\natexlab{b}}).

\bibitem[{\citenamefont{Chandrasekhar}(1943)}]{chandra}
\bibinfo{author}{\bibfnamefont{S.}~\bibnamefont{Chandrasekhar}},
  \bibinfo{journal}{Rev. Mod. Phys.} \textbf{\bibinfo{volume}{15}},
  \bibinfo{pages}{1} (\bibinfo{year}{1943}).

\bibitem[{\citenamefont{Seinfeld and Pandis}(1997)}]{seinfeld}
\bibinfo{author}{\bibfnamefont{J.~H.} \bibnamefont{Seinfeld}} \bibnamefont{and}
  \bibinfo{author}{\bibfnamefont{S.~N.} \bibnamefont{Pandis}},
  \emph{\bibinfo{title}{Atmospheric chemistry and physics: Air pollution to
  climate change}} (\bibinfo{publisher}{John Wiley and Sons},
  \bibinfo{year}{1997}).

\bibitem[{\citenamefont{Wang and Uhlenbeck}(1945)}]{wang}
\bibinfo{author}{\bibfnamefont{M.~C.} \bibnamefont{Wang}} \bibnamefont{and}
  \bibinfo{author}{\bibfnamefont{G.~E.} \bibnamefont{Uhlenbeck}},
  \bibinfo{journal}{Rev. Mod. Phys.} \textbf{\bibinfo{volume}{17}},
  \bibinfo{pages}{323} (\bibinfo{year}{1945}).

\bibitem[{\citenamefont{Hopkins et~al.}(2004)\citenamefont{Hopkins, Mitchem,
  Ward, and Reid}}]{hopkins}
\bibinfo{author}{\bibfnamefont{R.~J.} \bibnamefont{Hopkins}},
  \bibinfo{author}{\bibfnamefont{L.}~\bibnamefont{Mitchem}},
  \bibinfo{author}{\bibfnamefont{A.~D.} \bibnamefont{Ward}}, \bibnamefont{and}
  \bibinfo{author}{\bibfnamefont{J.~P.} \bibnamefont{Reid}},
  \bibinfo{journal}{Phys. Chem. Chem. Phys.} \textbf{\bibinfo{volume}{6}},
  \bibinfo{pages}{4924} (\bibinfo{year}{2004}).

\bibitem[{\citenamefont{Burnham and McGloin}(2006)}]{mcgloin}
\bibinfo{author}{\bibfnamefont{D.~R.} \bibnamefont{Burnham}} \bibnamefont{and}
  \bibinfo{author}{\bibfnamefont{D.}~\bibnamefont{McGloin}},
  \bibinfo{journal}{Opt. Express.} \textbf{\bibinfo{volume}{14}}
  (\bibinfo{year}{2006}).

\bibitem[{\citenamefont{Gittes and Schmidt}(1998)}]{gittes}
\bibinfo{author}{\bibfnamefont{F.}~\bibnamefont{Gittes}} \bibnamefont{and}
  \bibinfo{author}{\bibfnamefont{C.}~\bibnamefont{Schmidt}},
  \bibinfo{journal}{Opt. Lett.} \textbf{\bibinfo{volume}{23}},
  \bibinfo{pages}{7} (\bibinfo{year}{1998}).

\bibitem[{\citenamefont{O'Neil and Padgett}(2001)}]{oneill}
\bibinfo{author}{\bibfnamefont{A.~T.} \bibnamefont{O'Neil}} \bibnamefont{and}
  \bibinfo{author}{\bibfnamefont{M.~J.} \bibnamefont{Padgett}},
  \bibinfo{journal}{Opt. Commun} \textbf{\bibinfo{volume}{193}},
  \bibinfo{pages}{45} (\bibinfo{year}{2001}).

\end{thebibliography}

\end{document}